# Electronic and nuclear dynamics for low attosecond resolution of electronic chirality of Iodoacetylene

**Tianlv Xu[1], Jiawen Kong[1], Tianjing Zhou[1], Yan Wang[1], Jingqin Tu[1], Alireza Azizi[2], Herbert Früchtl[3], Tanja van Mourik[3], Steven R. Kirk[1] and Samantha Jenkins[*1]**

[1]*Key Laboratory of Chemical Biology and Traditional Chinese Medicine Research and Key Laboratory of Resource National and Local Joint Engineering Laboratory for New Petro-chemical Materials and Fine Utilization of Resources, College of Chemistry and Chemical Engineering, Hunan Normal University, Changsha, Hunan, 410081, China.*
[2]*State Key Laboratory of Powder Metallurgy, School of Materials Science & Engineering, Central South University, Changsha, Hunan, 410083, China*
[3]*EaStCHEM School of Chemistry, University of St Andrews, North Haugh, St Andrews, Fife KY16 9ST, Scotland, United Kingdom*

*To whom correspondence should be addressed: samanthajsuman@gmail.com

Attosecond science is an emerging topic where chirality plays a central role. Here we demonstrate subjecting iodoacetylene, a geometrically achiral molecule, to a pair of simulated non-ionizing ultrafast circularly polarized laser pulses at the highest time resolution to date, by two orders of magnitude (3.87 attoseconds), of the continuously-valued **S** and **R** electronic chirality assignments. We partner the only vector-based quantum chemical physics theory enabling full symmetry-breaking with electronic and nuclear dynamics simulations: the former does not require charge density differences or special symmetry positions. The resulting 'easy' and 'hard' directions of the total electronic charge density motion are quantified as a cardioid-like morphology for the duration of the simulated laser pulses and toroidal afterwards. Future research directions include determination of the underlying mechanism governing chiral induced spin selectivity, in addition to application to chiral spin selective phenomena in opto-spintronics and exotic superconductors, partnered with orbital-free density functional theory (OF-DFT).

## 1. Introduction

Attosecond science[1] is expanding across a wide range of scientific and technological applications including computing and medicine. There is, however, a long-standing lack of theoretical methods which scale well with the number of atoms N, specifically for ultrafast processes on the attosecond (as) timescale, required for understanding chiral effects of ultrafast circularly polarized laser pulses on electronic and nuclear dynamics. Traditional theories used for attosecond science are based on scalar quantities, e.g. orbital occupations, energies and atomic geometries and consequently do not allow progress beyond basic visualizations of electron or nuclear dynamics, employing e.g. differences in relative energies, charge densities and/or the use of special symmetry positions.

Conventional views consider chirality as a static structural property, in terms of the 'handedness' of molecular geometries, illustrated by the inability to superimpose left- and right-handed mirror-image structures. Using ultrafast circularly polarized laser pulses, it has been demonstrated that electrons can move in chiral paths on the attosecond timescale within molecules and materials[2] and hence are critical to understanding chiral phenomena[1,3–7]. Later it was realized that the missing helical behavior required to explain chirality was not discoverable by geometric or steric effects but by details of the electronic charge density distribution[8,9] for example, in the helical electronic transitions of spiroconjugated (cumulene) molecules[10,11]. Limitations of *all* scalar methods include the inability to differentiate between phenomena where relative energies are identical, such as the *S* and *R* enantiomers of a geometrically chiral molecule. *Full* symmetry-breaking, not to be confused with point-group or space-group symmetry operations, constitutes the removal of the dependency on measurable differences in atomic geometries and the corresponding relative energies and also doesn't require use of charge density differences or special symmetry positions.

Full symmetry-breaking is implemented within next generation quantum theory of atoms in molecules (NG-QTAIM)[12–14]. NG-QTAIM is the *vector*-based (directional) version of the original scalar QTAIM theory[15]. QTAIM is based on the spatial curvature of the total electronic charge density $\rho(\mathbf{r})$ at critical points, defined as locations where the gradient $\nabla\rho(\mathbf{r}) = 0$, in particular *bond* critical points (*BCP*s) at the saddle-point of $\rho(\mathbf{r})$ between bonded atoms, see the **Supplementary Materials S1** for the background of original (scalar) QTAIM. NG-QTAIM was previously used to distinguish the S and R geometric enantiomers of lactic acid[16], chiral $S_N2$ reactions, including intermediates[17] and alanine subjected to an electric field[18], without using the Cahn–Ingold–Prelog (CIP) priority rules to determine the chiral configuration (R/S)[19,20]. Another implication of full symmetry-breaking of the electronic and nuclear dynamics is that the electronic chirality will be a continuous, not binary, variable. NG-QTAIM has also been used for examination of ultrafast phenomena: in particular, irradiation by simulated non-ionizing circularly polarized laser pulses[21,22] which uncovered a geometric Berry phase that arises in an exotic superconductor[23].

Attosecond charge migration properties are well studied theoretically[24–27] and experimentally[28], along with theoretical investigations of chirality reversal[29]. The iodoacetylene molecule and cation have previously been examined with experiment[30] and theory[26]. The iodoacetylene molecule provides a bridge between complex real-world applications and simple theoretical models. The linear geometry of iodoacetylene enables direct probing of the electron dynamics in the presence of nuclear dynamics to capture the chirality induced by circularly polarized laser pulses, without the complicating factor of the short decoherence times (≈10-15 femtoseconds (fs)) of the iodoacetylene cation[26]. Additionally, we consider neutral iodoacetylene since neutral molecules[31] have been examined by emerging experimental techniques that do not require ionization such as photoexcitation circular dichroism experiments using non-ionizing ultrafast laser pulses[32] as well as nonlinear optics[33]. We demonstrate the attosecond variation of the magnitudes of the NG-QTAIM **S** and **R** electronic chirality with a pair of simulated clockwise {CW, [+1]}, right-handed (**R**) and counter-clockwise {CCW, [-1]}, left-handed (**S**) circularly polarized non-ionizing ultrafast laser pulses, see **Scheme 1** and the **computational details**. Note that NG-QTAIM **R** and **S** bond-chirality assignments are denoted by a bold font.

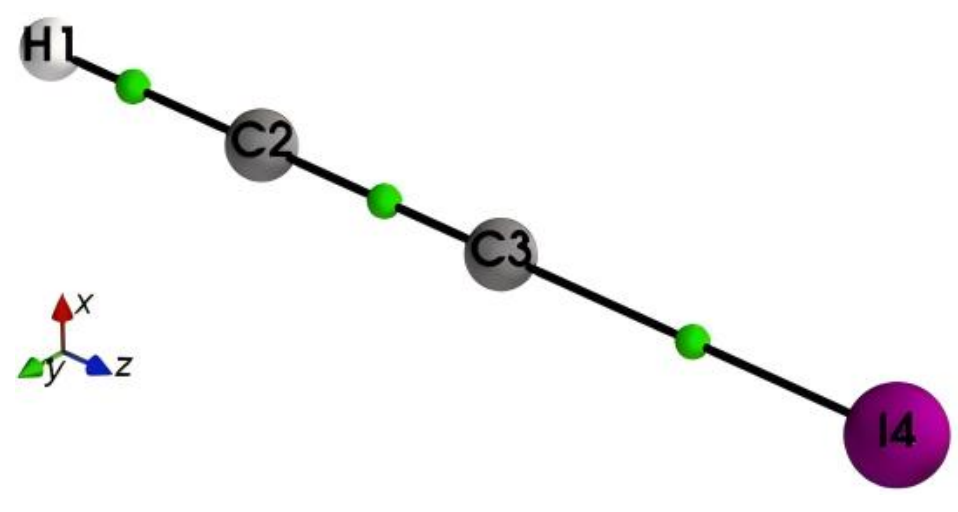

**Scheme 1**. **The iodoacetylene molecular graph**. A pair of simulated clockwise {CW, [+1]}, right-handed and counter-clockwise {CCW, [-1]}, left-handed circularly polarized non-ionizing ultrafast laser pulses will be applied along the ***z*** Cartesian coordinate axis, with the plane of polarization being the ***xy*** plane, notice the inset ***x-y-z*** Cartesian axes. The bond critical points (*BCP*s) are represented by the undecorated green spheres: these are extracted using our QuantVec program 'critic2sumviz', see the computational details.

## 2. Theoretical background and Computational Details

*The metallicity ξ(r_b) and the eigenvector projection time series*

In this work the C3-I4 *BCP* is a closed-shell *BCP*, the H1-C2 *BCP* and C2-C3 *BCP* are shared-shell *BCP*s. The Laplacian $\nabla^2\rho(\mathbf{r_b})$, where $\mathbf{r_b}$ indicates the position of the *BCP*, is positive and negative at a closed-shell *BCP* and shared-shell *BCP* respectively, see **Scheme 1**. An earlier observation for the *relative* tendency of the total electronic charge density $\rho(\mathbf{r}_b)$ to remain at a closed-shell *BCP* compared with motion towards the QTAIM atomic basins, resulting in lower *positive* values of the Laplacian $\nabla^2\rho(\mathbf{r_b})$, led to the origin of the metallicity $\xi(\mathbf{r_b})$[34]. This relative tendency for the $\rho(\mathbf{r_b})$ to move away from a closed-shell *BCP* is defined as the metallicity $\xi(\mathbf{r_b})$:

$$\xi(\mathbf{r_b}) = \rho(\mathbf{r}_b)/\nabla^2\rho(\mathbf{r}_b) \text{ for } \nabla^2\rho(\mathbf{r}_b) > 0 \quad \mathbf{(1)}$$

Values of the metallicity $\xi(\mathbf{r_b}) < 1$ from equation (**1**) for closed-shell *BCP*s correspond to non-metallic or insulating *BCP*s: conversely, values of $\xi(\mathbf{r_b}) \geq 1$ correspond to the presence of metallic *BCP*s. The metallicity $\xi(\mathbf{r}_b)$ relates inversely to "nearsightedness" of the first-order density matrix and also to the relative delocalization of the total electronic charge density $\rho(\mathbf{r}_b)$[35].

The Hessian of the total electronic charge density $\rho(\mathbf{r})$ eigenvector $\pm\underline{\mathbf{e}}_2$ corresponds to the 'easy' *direction* for *BCP* electrons to be displaced when the *BCP* is subjected to a displacement[36]. Conversely, the $\pm\underline{\mathbf{e}}_1$ eigenvector corresponds to the 'hard' direction for *BCP* electrons displacement, see the **Supplementary Materials S1**. We define a time-ordered set of points whose sequence is described by the parameter *s*, allowing the definition of *BCP* shift vector $\mathbf{dr} = \mathbf{r}_b(s) - \mathbf{r}_b(s\text{-}1)$, between each successive (in time) pair of molecular graphs. The Hessian of $\rho(\mathbf{r})$ eigenvectors $\{\underline{\mathbf{e}}_1, \underline{\mathbf{e}}_2, \underline{\mathbf{e}}_3\}$ were extracted for the first molecular graph in the sequence for each *BCP* and used for all subsequent *BCP* shift vectors $\mathbf{dr}$. We calculated scalar dot products to form the eigenvector projections $t_1$, $t_2$ and $t_3$, not to be confused with the time *t*, using equation **(2)**:

$$t_1 = \underline{\mathbf{e}}_1.\mathbf{dr},\ t_2 = \underline{\mathbf{e}}_2.\mathbf{dr},\ t_3 = \underline{\mathbf{e}}_3.\mathbf{dr} \qquad \textbf{(2)}$$

for each successive molecular graph in the sequence using the QuantVec 'drproject3' program, see the **Figure S4(a-i)** of the **Supplementary Materials S4** and the computational details.

*Generation of full symmetry-breaking for electronic and nuclear dynamics*

Within NG-QTAIM, each bond critical point (*BCP*) shift vector $\mathbf{dr}$ is mapped to a point, see equation **(2)**, in the construction of the components of the eigenvector-space trajectory $\mathbf{T}(s)$, denoted by a bold font. The mapping of the *BCP* shift vector $\mathbf{dr}$ to a point is responsible for full symmetry-breaking that can distinguish enantiomers at equilibrium configuration with degenerate, i.e. identical, energies, without *requiring* the presence of measurable differences in relative atomic positions.

The $\mathbf{T}(s)$ quantifies the (**F**) bond-flexing ('hard', $\pm\underline{\mathbf{e}}_1$), (**C**) bond-chirality ('easy', $\pm\underline{\mathbf{e}}_2$) and (**A**) bond-axiality ($\pm\underline{\mathbf{e}}_3$) directions of the total electronic charge density $\rho(\mathbf{r}_b)$ motion, see equation **(3)**, equation **(4)** and equation **(5)** respectively. The Hessian of $\rho(\mathbf{r}_b)$ eigenvector-space trajectories $\mathbf{T}(s)$ of the total electronic charge density $\rho(\mathbf{r}_b)$ and the eigenvector-space distortion set {**F**, **C**, **A**}, all denoted by a bold font, were constructed using the *BCP* shift vectors $\mathbf{dr}$ induced by the ultrafast laser pulses. This was enabled by the construction of so-called T(*s*) 'bounding-boxes', firstly for the duration of the laser pulse ($0 \leq t \leq 2.00$ fs) and secondly for the after-pulse ($2.00$ fs $\leq t \leq 23$ fs) for each of the {**F**, **C**, **A**}, see **Figure 1**, **Table 1** and the **Supplementary Materials S6** that was used to construct **Table 1**. The bond-flexing **F** is defined as the difference between a pair of $\mathbf{T}(s)$ bounding-boxes that contain the maximum extent of the eigenvector projections $(\underline{\mathbf{e}}_1\cdot\mathbf{dr})_{\mathbf{max}}$ between the $\mathbf{T}(s)$ for the left-handed circularly polarized (**S**, CCW, [-1]) laser pulse and the $\mathbf{T}(s)$ for the right-handed circularly polarized (**R**, CW, [+1]) laser pulse:

$$\mathbf{F} = [(\underline{\mathbf{e}}_1\cdot\mathbf{dr})_{\mathbf{max}}]_{\mathbf{CCW}} - [(\underline{\mathbf{e}}_1\cdot\mathbf{dr})_{\mathbf{max}}]_{\mathbf{CW}} \qquad \textbf{(3)}$$

and equation **(3)** provides a measure of the 'flexing-strain' of a bond-path associated with the $\pm\underline{\mathbf{e}}_1$ eigenvector when subject to an external force e.g. the electric field from a pair of simulated laser pulses.

The bond-chirality **C** is correspondingly defined as:

$$\mathbf{C} = [(\underline{\mathbf{e}}_2\cdot\mathbf{dr})_{max}]_{CCW} - [(\underline{\mathbf{e}}_2\cdot\mathbf{dr})_{max}]_{CW} \quad \mathbf{(4)}$$

The bond-chirality **C** in equation **(4)** quantifies the *BCP* 'bond-twist' distortion around the bond-path associated with the $\pm\underline{\mathbf{e}}_2$ eigenvector. The sign (+) or (-) of the continuously-valued bond-chirality **C** determines the **S** assignment for $\mathbf{C} > 0$ and **R** assignment for $\mathbf{C} < 0$, see **Table 1**.

The bond-axiality **A** is correspondingly defined as:

$$\mathbf{A} = [(\underline{\mathbf{e}}_3\cdot\mathbf{dr})_{max}]_{CCW} - [(\underline{\mathbf{e}}_3\cdot\mathbf{dr})_{max}]_{CW} \quad \mathbf{(5)}$$

The bond-axiality **A** in equation **(5)** quantifies the direction of *axial* displacements of *BCP*s in response to the (**S**, CCW, [-1]) and (**R**, CW, [+1]) circularly polarized ultrafast laser pulses, i.e. the gliding of the *BCP* along the bond-path associated with the $\pm\underline{\mathbf{e}}_3$ eigenvector[37].

The chirality-helicity function $\mathbf{C}_{helicity} = |\mathbf{C}|\mathbf{A}$, is defined as the numerical product of the bond-chirality **C** and bond-axiality **A** and was earlier used to quantify the nature of any helical character of the *BCP* motion due to the effect of a pair of simulated left- and right-circularly polarized ultrafast laser pulses[10]. The chirality-helicity function $\mathbf{C}_{helicity}$ is consistent with non-ionizing photoexcitation circular dichroism experiments on neutral chiral molecules that utilized the helical motion of the bound electrons for chiral discrimination[32]. The Kolmogorov-Zurbenko[38] data filter was applied as the smoothing procedure for the calculation of the $\mathbf{T}(s)$, see the **Supplementary Materials S2**.

In this first investigation using NG-QTAIM with both nuclear and electron dynamics, we use simulated non-ionizing circularly polarized ultrafast laser pulses of duration 2.0 fs to induce the electron and nuclear dynamics of neutral iodoacetylene for an additional 23 fs after the laser pulses. The frozen nuclei approximation was not used because of significant bond-stretching of the H1-C2, C2-C3 and C3-I4 bonds quantified by the bond-stretch values after the laser pulses were removed, see **Figure S5(a-c)** of the **Supplementary Materials S5**. In particular, the C3-I4 bond-stretch is large, as expected due to the much larger polarizability of iodine compared to carbon. All *BCP* ellipticity ε values were insignificant for both the during-laser pulse and after-laser pulse time intervals, see **Figure S5(d-f)**. The large C3-I4 bond-stretch values, see the **Figure S5(c)**, are consistent with the very large values of the QTAIM metallicity $\xi(\mathbf{r}_b)$, with values of metallicity $\xi(\mathbf{r}_b) \geq 3$, explaining the ease of motion of the C3-I4 *BCP* and the C3 and I4 nuclei, see **Figure S5(i)**. The bond-stretch, ellipticity ε and metallicity $\xi(\mathbf{r}_b)$ are calculated using the QuantVec program ‘cpscalar’, see the computational details.

**3. Computational Details.** *The model: Iodoacetylene electron and nuclear dynamics driven by circularly polarized laser pulses Theory level and simulation box.* The initial structure of the ground state of the iodoacetylene molecule, optimized at the CCSD(T)/cc-pVQZ/ECP28MDF level of theory, was taken from literature[30]. We used orbital-free DFT (OF-DFT), with its advantages[14] of near-linear scaling for large numbers of atoms $N$, total charge density-based $\rho(\mathbf{r})$ protocols and machine learning-based DFT kinetic energy density functionals as implemented in the Octopus code[39] (development version ‘Chierchiae’), which defines wavefunctions on a real-space grid, for the electronic and nuclear dynamics calculations. The Octopus input files are provided in the **Supplementary Materials S3**.

The simulation box for the real-space calculations was defined as a cuboid aligned along the ***x-y-z*** orthogonal Cartesian axes, with dimensions 8.0Å × 8.0Å ×18.0Å respectively. The centre of the box was placed at the Cartesian axes origin. The initial linear molecular structure was then aligned on the ***z***-axis at the centre of the simulation box, see **Scheme 1**. The size of the simulation box was chosen as large enough, later confirmed after initial total electronic charge density $\rho(\mathbf{r})$ calculation, to ensure that the $\rho(\mathbf{r})$ at the edges of the box and hence potential boundary wave-packet reflection effects are negligible. The mesh spacing for the simulation box real-space grid was selected to be 0.05Å, a spacing known from previous work to provide well-converged QTAIM and NG-QTAIM properties[23,40].

*Electronic structure theory and geometry optimization.* All electronic structure calculations were undertaken at the orbital-free LDA DFT (OF-DFT) level, using Hartwigsen-Goedecker-Hutter[41] pseudopotentials and the efficient ‘chebyshev_filter’ eigensolver as implemented in the Octopus code[39], with an eigensolver tolerance of $10^{-8}$ and a convergence criterion on the relative density change in the SCF iterations of $10^{-9}$, using a Broyden mixing scheme based on the potential. All wavefunctions were computed for the ‘unpolarized’ total electronic charge density $\rho(\mathbf{r})$. The ‘FIRE’ geometry optimization algorithm implemented in the Octopus code[42] was used and the molecular structure converged to a force threshold of 0.01 eV/Å. The resulting optimized linear structure was then aligned along the ***z***-axis with ***x*** = 0, ***y*** = 0, with the centre of mass of the molecule located at the ***x-y-z*** Cartesian axes origin, i.e. the geometric centre of the simulation box.

*Ground state and excited state preparation, electronic and nuclear dynamics propagation.* The initial electronic ground state was computed for the geometry-optimized structure using the aforementioned electronic convergence parameters. In addition, the 8 lowest-energy ‘unoccupied’ excited states were also converged using the same convergence criteria, in preparation for subsequent laser pulse excitation. The nuclei were treated as particles and Ehrenfest dynamics was used to perform the electronic and nuclear time propagation. A propagation timestep of 0.01 au (0.242 attoseconds) was selected, ensuring both numerical stability and computational efficiency, with the total energy being updated at every propagation timestep. Both electrons and nuclei were permitted to move during the propagation. The enforced time-reversal symmetry (ETRS) propagator implemented in the Octopus code was used[43] with a Lanczos exponential method[44] and time propagation self-consistency was enforced at every propagation timestep, given the non-linear nature of the time-dependent Kohn-Sham equations. Additionally, the electronic ground state was recomputed, using SCF cycles, every 50 propagation timesteps, maintaining electronic consistency as the nuclei moved. The total electronic charge density $\rho(\mathbf{r})$ grid, in Gaussian .cube format and the number of excited electrons were calculated using the regularly updated ground state. The $\rho(\mathbf{r})$ along with the molecular geometry in conventional .xyz format and the components of the imposed laser electric field $\mathbf{E}(t)$ were saved every 16 propagation timesteps.

*Laser pulse parameters.* The applied laser pulse $\mathbf{E}(t)$ was simulated using time-dependent electric field components [$E_x(t)$, $E_y(t)$] in the ***xy*** polarization plane, see the molecular coordinate axes as defined in **Scheme 1**. These electric field components [$E_x(t)$, $E_y(t)$] were specified as the product of the following three terms:

- The 'carrier' wave of unit amplitude, starting at time $t = 0$ fs defined to be the vector [$\sin(\omega t)$, $\sin(\omega t + \pi/2)$] in the ***xy*** polarization plane with selected laser pulse excitation frequency $\omega$.
- An 'envelope' of amplitude corresponding to a single laser pulse as a half-cycle 'sin-squared' function $E\sin^2(\omega_{env}t)$, also starting at time $t = 0$ with phase = 0° and peak electric field amplitude E = 100 x $10^{-4}$ a.u. with $\omega_{env}$ selected to result in a half-cycle, i.e. single laser pulse, duration of 2.0 fs.
- A 'polarization' $p$ factor of [+1] or [-1], applied uniquely to the $E_y$-field component. This yielded an electric field $\mathbf{E}(t)$ vector which turned either clockwise {CW, [+1]} and right-handed (**R**) or counter-clockwise {CCW, [-1]} and left-handed (**S**) in the ***xy*** polarization plane.

Combining these three factors yields equation **(6)** for the electric field $\mathbf{E}(t)$ vector of the laser pulse:

$$\mathbf{E}(t) = [E_x(t), E_y(t)] = E\sin^2(\omega_{env}t)[\sin(\omega t), p\sin(\omega t + \pi/2)]; \text{ where } p = [+1] \text{ or } p = [-1] \qquad \mathbf{(6)}$$

Given the short (2.0 fs) duration of the laser pulses, a 'field broadening' approach to constructing the time-evolving superposition of ground and excited electronic states was employed: in the special case of a '$\sin^2$' amplitude envelope function laser pulse of time duration $\Delta T$ the energetic 'broadening' induced by the field $\Delta E = \hbar/\Delta T$. A carrier frequency i.e., excitation energy, $\omega$ = 5.60 eV, corresponding to the first excited state energy, was therefore chosen for the ultrafast laser pulse, the energetic broadening covering a range of possible lower-lying excited states. This choice is sufficient to create a superposition of the ground state and one or more excited states as demonstrated by the computed number of excited electrons, see **Figure S5(j)** of the **Supplementary Materials S5**.

*Extraction of QTAIM and NG-QTAIM quantities*

We performed the NG-QTAIM calculations using our QuantVec[45] code. From the computed time sequence of molecular graphs, the following scalar QTAIM quantities were evaluated for each molecular graph using the QuantVec program 'cpscalar': the total electronic charge density $\rho(\mathbf{r_b})$, ellipticity ε, metallicity $\xi(\mathbf{r_b})$ and the shortest separation geometric distance between bonded nuclei (GBL) for each *BCP*. For each of the time-ordered total electronic charge density grids, the CRITIC2[46] code was used to extract the bond critical points (*BCP*s) from $\rho(\mathbf{r})$. As the Octopus[39] calculation used HGH pseudopotentials, core electronic charge densities were added to recreate the full total electronic charge density $\rho(\mathbf{r})$ distribution in each grid. The so-called 'density smoothing' algorithm in CRITIC2[47] was used with 2030 seed points distributed in the simulation cell using a level 2 Wigner-Seitz cell subdivision, with a nuclear attractor capture radius of 0.29 au and spurious QTAIM topological features removed if their $\rho(\mathbf{r})$ fell below $10^{-5}$ au. The Poincaré-Hopf relationship was satisfied throughout, see the **Supplementary Materials S1**. The CRITIC2 output for each propagation timestep was converted to a .sumviz-format molecular graph file containing data on the *BCP*s using the QuantVec program 'critic2sumviz'. The QuantVec 'drproject3' program uses the Hessian of $\rho(\mathbf{r_b})$ eigenvectors $\{\underline{\mathbf{e}}_1, \underline{\mathbf{e}}_2, \underline{\mathbf{e}}_3\}$ extracted from the *first* molecular graph to construct equation **(2)** and **Figure 1** and from each *successive* molecular graph to construct **Figure 2**. The QuantVec 'trajplot' program calculates the bounding-boxes of the $\mathbf{T}(s)$ that were used to assemble the $\{\mathbf{F}, \mathbf{C}, \mathbf{A}\}$, see equations **(3-5)** and **Table 1**. See the **Supplementary Materials S2** for details on calculation of the NG-QTAIM trajectories $\mathbf{T}(s)$.

## 4. Results and Discussion

For the 2.0 fs duration of the laser pulses the C2-C3 *BCP* and C3-I4 *BCP* possess very significant degrees of electronic bond-chirality **C**, see **Table 1**. This can be seen by comparison of the bond-chirality **C** = -0.0044 (a triple bond) of the C2-C3 *BCP* presented here with earlier work on the C-C *BCP* (a double bond) of ethane (**C** = 0.0026), also subject to a pair of simulated circularly polarized lasers but with a larger peak E-field, E = 200.0 x $10^{-4}$ au[21]. The value of the bond-flexing **F** = 0.0091 for the C3-I4 *BCP* is rather large, reflecting the highest C3-I4 bond deformation due to the laser pulses along with the very high $\xi(\mathbf{r_b})$ metallicity values ($\xi(\mathbf{r_b}) \geq 3.0$), see the **Figure S5(i)** of the **Supplementary Materials S5**, consistent with the much larger polarizability of iodine compared to carbon.

After the laser pulses are removed there is a significant reduction in the values of the bond-chirality **C** for the C2-C3 *BCP* and C3-I4 *BCP*. None of the values of the chirality-helicity function $\mathbf{C}_{helicity} = |\mathbf{C}|\mathbf{A}$, however, are significant for both the duration of the laser pulses and the after-pulses since all are below 1.0 x $10^{-5}$. This indicates a lack of the helical motion of the electrons associated with the *BCP*s which is only expected for a geometrically chiral molecule, i.e. one that possesses molecular enantiomers[18].

**Table 1. The iodoacetylene eigenvector-space *BCP* T(s) distortion sets {F, C, A} during and after the laser pulses.** The eigenvector-space distortion set {**F**, **C**, **A**}, where **F** = bond-flexing, **C** = bond-chirality and **A** = bond-axiality for the bond critical point (*BCP*) eigenvector-space trajectories **T**(s) with the application of a pair of right-handed clockwise {CW, [+1]} and left-handed counter-clockwise {CCW, [-1]} simulated non-ionizing circularly polarized laser pulses of duration 2.0 fs with peak electric (E)-field E = 100.0 x $10^{-4}$ a.u in the ***xy*** plane. The NG-QTAIM electronic chirality **C** assignments **S** or **R** are provided in the square brackets. Results are presented for the duration of the laser pulses ($0 \leq t \leq 2.00$ fs) and of the after-pulses (2.00 fs $\leq t \leq$ 23 fs). The QuantVec 'trajplot' program is used to construct the {**F**, **C**, **A**}.

| **Eigenvector-space *BCP* T**(s) | {**F**, **C**, **A**} |
| --- | --- |
| *During-pulse* | |
| *H1-C2* | {0.001770, 0.000290**[S]**, -0.000070} |
| *C2-C3* | {0.001601, -0.004449**[R]**, 0.000241} |
| *C3-I4* | {0.009145, -0.005295**[R]**, 0.000251} |
| *After-pulse* | |
| *H1-C2* | {0.001042, 0.001305**[S]**, 0.000537} |
| *C2-C3* | {-0.002156, 0.000823**[S]**, -0.000360} |
| *C3-I4* | {-0.001197, 0.001941**[S]**, -0.000256} |

For the duration of the laser pulses all the **T**(*s*) possess a cardioid-like morphology when viewed in the bond-flexing **F** and bond-chirality **C** plane, see **Figure 1(a-c)**. Furthermore, the orientation of the cardioid-like morphology of the **T**(*s*) reflects the NG-QTAIM bond-chirality **C** assignment, during the laser pulses, which is **S** for the H1-C2 *BCP* **T**(*s*) and **R** for both the C2-C3 *BCP* **T**(*s*) and C3-I4 *BCP* **T**(*s*), see **Table 1**, as indicated by the relative orientations of the start points (pink/cyan spheres) of the **T**(*s*). The cardioid-like morphology, however, is lost after the laser pulses are removed ($t > 2.0$ fs) where the **T**(*s*) all possess a toroidal morphology, see **Figure 1(d-f)**.

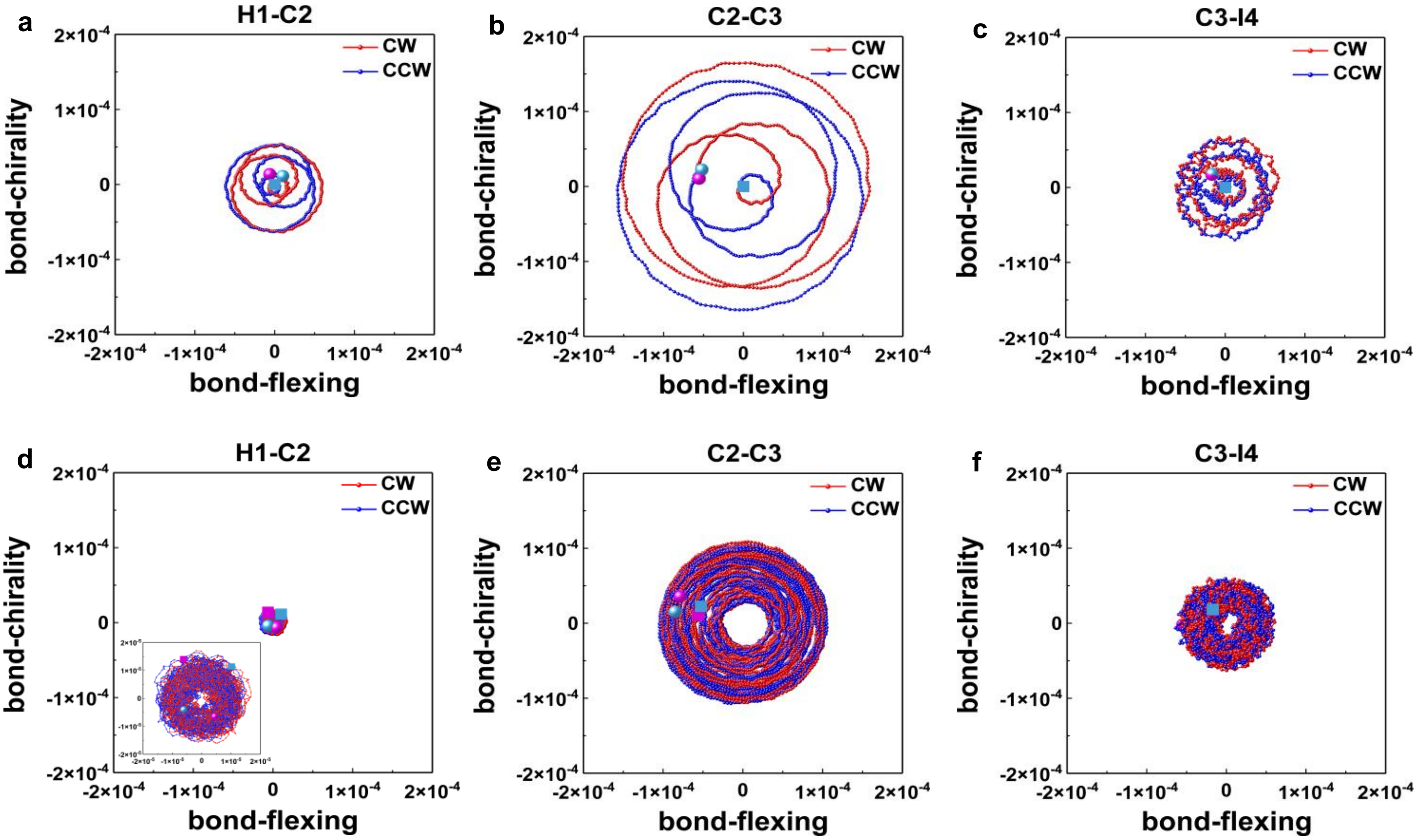

**Figure 1**. **Full symmetry-breaking in eigenvector-space using CW and CCW circularly polarized laser pulses.** View down the bond-axiality **A** axis of the Hessian of $\rho(\mathbf{r})$ eigenvector-space trajectories $\mathbf{T}(s)$ for each bond critical point (*BCP*) of the iodoacetylene molecular graph. Results are presented during ($0 \leq t \leq 2.00$ fs) the simulated non-ionizing circularly polarized laser pulses (sub-figures **(a-c)**) and after ($t > 2.0$ fs) the laser pulses (sub-figures **(d-f)**). Note, for comparison the same axes scales are used throughout. The inset of sub-figure **(d)** displays a 10x magnification to highlight the toroidal morphology. Start and end points of the $\mathbf{T}(s)$ are denoted for CW by the pink spheres and cubes and for CCW by cyan spheres and cubes respectively. The QuantVec 'drproject3' program was used to construct the $\mathbf{T}(s)$. For further details see the caption of **Table 1**.

The variation of the bond-chirality **C** assignments with time *t* in terms of the **S** (blue) and **R** (red) for each of the H1-C2 *BCP*, C2-C3 *BCP* and C3-I4 *BCP* are presented in **Figure 2(a-c)**, the components $t_2$ using equation **(2)** are provided in the **Supplementary Materials S1**. This was undertaken using the vector-based NG-QTAIM to quantify the continuous electronic **S** and **R** assignments using bond-chirality **C** on an attosecond resolution, as a consequence of irradiation by the non-ionizing circularly polarized laser pulses, see **Figure 2**.

Here for the work on iodoacetylene, with the simulated laser pulse polarization plane oriented perpendicular to the molecular axis[22], the value of the chirality-helicity function $\mathbf{C}_{helicity} \approx 0$ as expected due to iodoacetylene being formally achiral resulting from insignificant values of bond-axiality **A**. Conversely, for a geometrically chiral molecule the chirality-helicity function $\mathbf{C}_{helicity} \neq 0$ due to the higher asymmetry and presence of significant values of both the bond-axiality **A**[48] and bond-chirality **C**. Additionally, the cardioid-like morphology we see in this work for all three of the during-pulse *BCP* trajectories **T**($s$), see **Figure 1(a-c)**, has also been found using numerical simulations and experiments of fully chaotic billiard laser pulses[49]. For the duration of the pulses the C3-I4 *BCP* **T**($s$) possessed a somewhat 'jittery' morphology, see **Figure 1(c)**, despite the very low C3-I4 bond-stretch during the laser pulses, due to the very high C3-I4 *BCP* metallicity $\xi(\mathbf{r_b})$ for the entire simulation, see **Figure S5(c)** and **Figure S5(i)** of the **Supplementary Materials S5** respectively. After the laser pulses are removed the morphology of the H1-C2 *BCP* **T**($s$), C2-C3 *BCP* **T**($s$) and C3-I4 *BCP* **T**($s$) are all toroidal in form, see **Figure 1(d-f)**.

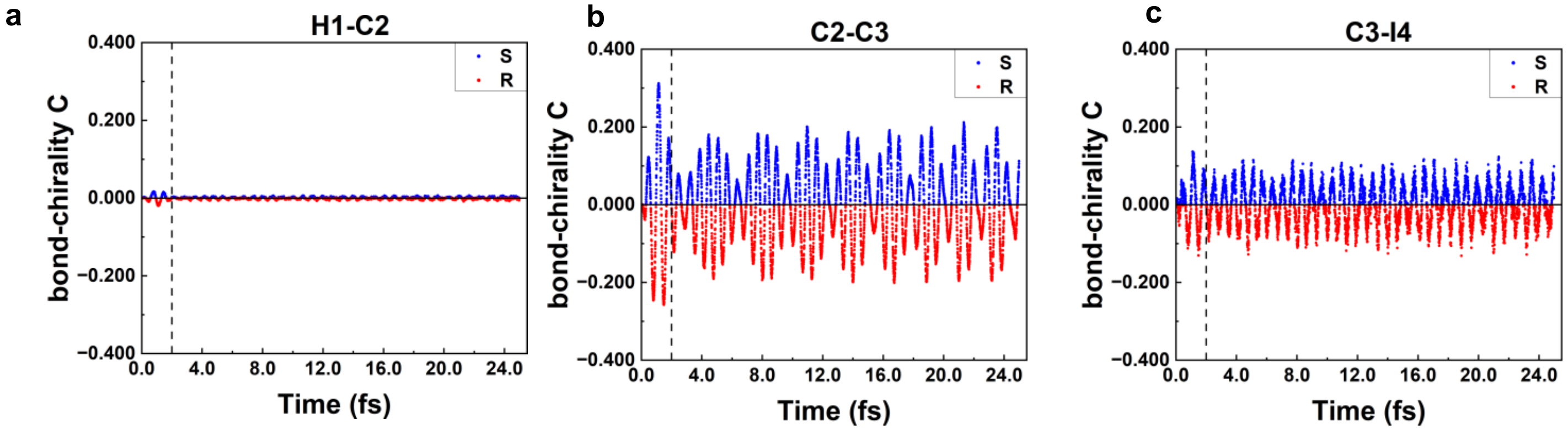

**Figure 2. Attosecond resolution time variation of the bond-chirality C of the iodoacetylene molecular graph**. The markers corresponding to the bond-chirality **C**, see equation **(4)**, assignments, **S** (blue) and **R** (red), of the bond critical point (*BCP*) eigenvector-space trajectories **T**($s$) are presented in sub-figures **(a-c)** respectively. Results are presented for the duration of the simulated non-ionizing circularly polarized laser pulses ($0 \leq t \leq 2.00$ fs) and the after-pulses ($t > 2.0$ fs), that are separated by the dashed vertical line at time $t$ = 2.00 fs. The time-resolution of the bond-chirality **C** is 3.87 (as) attoseconds. One full cycle of the laser pulse carrier period is 0.7392 femtoseconds. The QuantVec 'drproject3' program is used to calculate the **T**($s$). The QuantVec 'trajplot' program is used to calculate the values of the bond-chirality **C**.

The magnitude of the chirality **C** for the C2-C3 *BCP* in this investigation was consistent with the chirality **C** value of the C-C *BCP* in ethane, also subjected to non-ionizing circularly polarized lasers, but with a larger peak E-field and a double C-C bond, earlier obtained using GGA TD-DFT (orbital-based) methods and time dependent configuration interaction[21]. In addition, the values of the ellipticity ε were consistently negligible for both the CW and CCW laser pulses for all *BCP*s for the entirety of the simulations as expected for the linear iodoacetylene molecule. The corresponding CW and CCW values were indistinguishable for both the bond-stretch and metallicity $\xi(\mathbf{r_b})$, as expected for scalar measures. Note, the ellipticity ε has previously displayed *partial* symmetry-breaking[22] with respect to the effect of circularly polarized lasers pulses, although never full symmetry-breaking. These findings therefore justify the use of LDA within orbital-free DFT (OF-DFT)[14] in this investigation.

## 5. Conclusions

In this investigation we discovered continuous **S** and **R** electronic chirality assignment reversals or 'flips' for formally achiral iodoacetylene with a 3.87 attosecond time resolution, the highest to date, consistent with time-resolved photoelectron circular dichroism experiment[50]. This is due to removing the dependency on the two orders of magnitude slower collective movement of the electronic charge density $\rho$ across an entire molecule as previously undertaken using Grimmes CCM[29]. NG-QTAIM extracts the eigenvectors $\{\underline{\mathbf{e}}_1, \underline{\mathbf{e}}_2, \underline{\mathbf{e}}_3\}$ from the Hessian of the electronic charge density $\rho(\mathbf{r}_b)$ at a *BCP* calculated from the curvatures (second derivatives) of $\rho(\mathbf{r}_b)$. Consequently, this enables a much more rapid and therefore sensitive response to the evolving electron dynamics than the previous[29] use of the electronic charge density $\rho$. With this discovery we therefore uncovered the implications of full symmetry-breaking for electronic and nuclear dynamics without using charge density differences or special symmetry positions. Developments using NG-QTAIM partnered with simulated laser pulse irradiation for any achiral or chiral molecule or solid are also possible, e.g. we earlier located a geometric Berry phase[23], as well as the inclusion of spin-orbit coupling with spinor wavefunction representation and further development of an analysis of a probe laser pulse[40]. Spin-orbit coupling with spinors however, only provides *partial* symmetry-breaking through the mechanism of splitting the spin-up and spin-down energy bands (or levels). NG-QTAIM, however, doesn't *require* this mechanism to provide full symmetry-breaking. In both these cases of NG-QTAIM applied to solids, the distortion set $\{\mathbf{F}, \mathbf{C}, \mathbf{A}\}$ were determined for so-called *crystal graphs*, the solid analogue of the molecular graph. Charged species such as the iodoacetylene cation are possible to consider, with the additional implementation of spin-orbit coupling, to provide an alternate approach to understanding coherence and decoherence/recoherence phenomena[26]. Ongoing developments include implementation of the *spin* current density trajectory $\mathbf{T}_{\mathbf{Jj}}(s)$, the subscripts '$\mathbf{J}$' and '$\mathbf{j}$' indicating the current density ($\mathbf{J}$) and spin ($\mathbf{j}$) to enable the calculation of the chiral spin current density selectivity $\mathbf{P}_C$, within the NG-QTAIM interpretation.

The full symmetry-breaking of electron and nuclear dynamics when subjected to circularly polarized laser irradiation, only accessible through NG-QTAIM, is obtainable for large molecules/surfaces/solids with N $\approx 10^4$ to $10^6$, where N is the number of atoms, as currently partnered with OF-DFT which enables near-linear scaling with N. Applications include determining the, as yet unknown, underlying mechanism governing chiral-induced spin-selectivity (CISS)[51]. The CISS effect could be investigated e.g. for hybrid perovskites with interstitial structurally chiral organic molecules such as R- and S-methylbenzylamine (MBA) and could be 'deconstructed' by manipulation of the electronic spin chirality[52] of these interstitial chiral molecules. An example of manipulation of the electronic bond-chirality $\mathbf{C}$ would be to vary the carrier envelope phase (CEP) angle with a pair of simulated circularly polarized laser pulses[21]. This could induce electronic *achiral* character, by reducing the value of the chirality-helicity function $\mathbf{C}_{helicity} \approx 0$[23]. Therefore, we would gain insight into CISS by reducing CISS to a so-called achiral induced spin selectivity (AISS). Laser pulses would be used to induce and control the helical motion of *BCP*s, quantified by the spin chirality-helicity function $\mathbf{C}_{\mathbf{j}helicity}$[40], to understand the previously discovered spin-selective behaviors in opto-spintronics[53–55] and exotic superconductors[56]. Further manipulation of the spin chirality-helicity function $\mathbf{C}_{\mathbf{j}helicity}$ could be achieved by the use of variable elliptically polarized laser pulses or switching the laser pulse direction. NG-QTAIM will be used in these future investigations using GGA and hybrid functionals within orbital-free DFT (OF-DFT) as well as employing non-adiabatic methods beyond the Ehrenfest approximation.

Emerging experimental techniques include chiral spectroscopy by dynamical symmetry-breaking in high harmonic generation[57], ultrafast imaging of chiral dynamics using a non-linear enantio-sensitive optical response to synthetic chiral light[58] and ultrafast photoelectron imaging of attosecond electron dynamics[59,60]. Attosecond chiroptical spectroscopy may provide experimental insights into open questions, such as the origin of the chirality-induced spin selectivity (CISS) effect and the role of a molecule's geometrical handedness on electron spin dynamics. The interplay between structural, as well as electronic, chirality and spin is an emerging area with potential to advance solid-state chiral photonics and chiral spintronics[6]. Common features in opto-spintronics phenomena between semiconductors and exotic superconductors, such as the minor role contributed by phonons, will be utilized. This work provides a demonstration of the importance of full symmetry-breaking of electron and nuclear dynamics for understanding the phenomena of attosecond science[48].

## Acknowledgements

Thanks are given for useful discussions with Yasuteru Shigeta, Marco Nascimento, Martin J. Paterson, Graham Worth, Grant Hill and Vincent Ortiz.

## Funding Information

The Hunan Natural Science Foundation of China project is thanked, approval number: 2022JJ30029. The One Hundred Talents Foundation of Hunan Province is also acknowledged for the support of S.J. and S.R.K. The use of SHARCNET computing facilities is thanked through our sponsor Paul W. Ayers. Tianlv Xu thanks the Education Department of Hunan Province for support with an Outstanding Youth Project of Education Department of Hunan Province (award iD: 23B0053).

## Competing Interests

The authors declare no competing interests.

# SUPPLEMENTARY INFORMATION

# Electronic and nuclear dynamics for low attosecond resolution of electronic chirality of Iodoacetylene

**Tianlv Xu[1], Jiawen Kong[1], Tianjing Zhou[1], Yan Wang[1], Jingqin Tu[1], Alireza Azizi[2], Herbert Früchtl[3], Tanja van Mourik[3], Steven R. Kirk[1] and Samantha Jenkins[*1]**

*[1]Key Laboratory of Chemical Biology and Traditional Chinese Medicine Research and Key Laboratory of Resource National and Local Joint Engineering Laboratory for New Petro-chemical Materials and Fine Utilization of Resources, College of Chemistry and Chemical Engineering, Hunan Normal University, Changsha, Hunan, 410081, China.*
*[2]State Key Laboratory of Powder Metallurgy, School of Materials Science & Engineering, Central South University, Changsha, Hunan, 410083, China*
*[3]EaStCHEM School of Chemistry, University of St Andrews, North Haugh, St Andrews, Fife KY16 9ST, Scotland, United Kingdom*

*To whom correspondence should be addressed: samanthajsuman@gmail.com

**Supplementary Materials S1.** QTAIM theoretical background.

*The QTAIM bond critical point (BCP) ellipticity ε.* The topology sum rule for the four types of critical points found in a molecule is the Poincaré-Hopf relation:

$$n - b + r - c = 1 \qquad \textbf{(S1)}$$

Within QTAIM these four types of critical points are: a local maximum ($n$), two types of saddle point ($b$ and $r$) and a local minimum ($c$), $n$, $b$, $r$ and $c$ are the numbers of nuclear attractor critical points (*NACP*s), bond critical points (*BCP*s), ring critical points (*RCP*s) and cage critical (*CCP*s), where the gradient $\nabla\rho(\mathbf{r}) = 0$ except for the nuclear critical point (*NACP*) that is located at the position of the nucleus where $\nabla\rho(\mathbf{r}) \neq 0$ and is instead a cusp. The critical points are classified using equation **(S1)** and the notation ($R$, ω) where $R$ is the rank of the Hessian matrix, i.e., the number of distinct non-zero eigenvalues (in increasing order $\lambda_1$, $\lambda_2$, $\lambda_3$) and ω is the signature (the algebraic sum of the signs of the eigenvalues), with corresponding eigenvectors $\underline{\mathbf{e}}_1$, $\underline{\mathbf{e}}_2$ and $\underline{\mathbf{e}}_3$. In the limit that the forces on the nuclei are zero, an atomic interaction line, the line passing through a *BCP* and terminating on two nuclear attractors along which the total electronic charge density $\rho(\mathbf{r})$ is locally maximal with respect to nearby lines, becomes a bond-path. The collection of all four types of critical points and the associated bond-paths are known as the *molecular graph*. The ellipticity, ε, defined as $\varepsilon = |\lambda_1|/|\lambda_2| - 1$, quantifies the relative accumulation of the electronic charge density $\rho(\mathbf{r}_b)$ distribution in the two directions $\pm\underline{\mathbf{e}}_1$ and $\pm\underline{\mathbf{e}}_2$ that are perpendicular to the bond-path at a Bond Critical Point (*BCP*) with position $\mathbf{r}_b$.

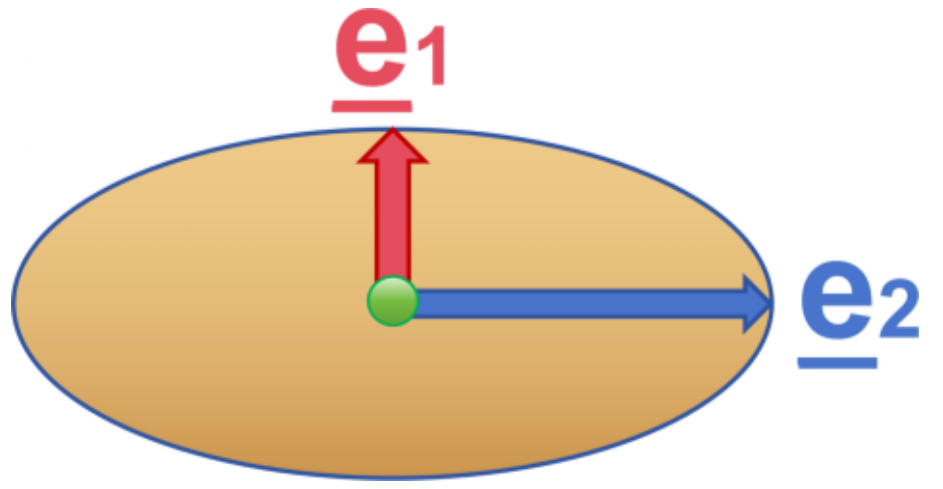

**Supplementary Scheme 1**. The cross section through a bond at the bond critical point (*BCP*), indicated by the green sphere. The $\lambda_1$ and $\lambda_2$ eigenvalues with associated eigenvectors $\pm\underline{\mathbf{e}}_1$ ('hard' direction) and $\pm\underline{\mathbf{e}}_2$ ('easy' direction) respectively, define the axes of the elliptical distribution of the total electronic charge density $\rho(\mathbf{r})$ in the vicinity of the bond critical point (*BCP*) and indicate the magnitudes of the least and greatest extents of the distribution of $\rho(\mathbf{r})$.

The ellipticity, ε, defined as $\varepsilon = |\lambda_1|/|\lambda_2| - 1$, quantifies the relative accumulation of the electronic charge density $\rho(\mathbf{r}_b)$ distribution in the directions $\pm\underline{\mathbf{e}}_1$ and $\pm\underline{\mathbf{e}}_2$ that are perpendicular to the bond-path at a Bond Critical Point (*BCP*) with position $\mathbf{r}_b$, see the **Supplementary Scheme 1**. For ellipticity $\varepsilon > 0$, the shortest and longest axes of the elliptical distribution of $\rho(\mathbf{r}_b)$ are associated with the $\lambda_1$ and $\lambda_2$ eigenvalues, respectively. From the electron-preceding perspective a change in the total electronic charge density $\rho(\mathbf{r})$ distribution that defines a chemical bond can cause changes in atomic positions.

**Supplementary Materials S2.** Procedures to generate the Eigenvector-Space Trajectory $\mathbf{T}(s)$

**2(a)** *Numerical Considerations for Calculation of the Eigenvector-Space Trajectory **T(s)**.* Central to the concept of the eigenvector-space trajectory $\mathbf{T}(s)$, denoted by a bold font, is the concept of a monotonically increasing sequence parameter $s$, which may take the form of an increasing integer sequence (0, 1, 2, 3,...) in applications where a set of discrete numbered steps are involved, or a continuous real number. The 3-D eigenvector-space trajectory $\mathbf{T}(s)$ is then defined as an ordered set of points, whose sequence is described by the parameter $s$. In this application, we used an integer step number for $s$, directly mapping to the timestep numbers in the electron and nuclear dynamics simulation at which the total electronic charge density $\rho(\mathbf{r})$ is saved. We first choose to associate $s = 0$ with a specific reference molecular graph, in this case, the initial molecular graph at the start of the electron and nuclear dynamics simulation. For a specific *BCP*, the eigenvector-space coordinates associated with each of the points in the sequence are calculated by evaluating the components of the *BCP* shift vector $\mathbf{dr} = \mathbf{r_b}(s) - \mathbf{r_b}(s\text{-}1)$ where $\mathbf{r_b}$ indicates the location of the *BCP*, from the previous step to the current step in the reference coordinate frame defined by the eigenvectors $\underline{\mathbf{e}}_1$, $\underline{\mathbf{e}}_2$, $\underline{\mathbf{e}}_3$. Hence each *BCP* shift vector $\mathbf{dr}$ is mapped to a point in eigenvector-space to form the eigenvector projections: [$t_1 = \underline{\mathbf{e}}_1.\mathbf{dr}$, $t_2 = \underline{\mathbf{e}}_2.\mathbf{dr}$, $t_3 = \underline{\mathbf{e}}_3.\mathbf{dr}$] and thus the sequence of these points in eigenvector-space comprises the 3-D eigenvector-space trajectory $\mathbf{T}(s)$.

The calculation of the $\mathbf{T}(s)$ is made easier if the code which produces the sequence of time-dependent total electronic charge density $\rho(\mathbf{r})$ distributions generates these structures at regularly-spaced points, in this case, every 16 electron and nuclear dynamics timesteps, see the main text. The consequence of this desirable characteristic is that there are few or no large changes or 'spikes' in the magnitude of the *BCP* shift vector $\mathbf{dr}$ i.e. $\Delta\mathbf{dr}$, between path step $\boldsymbol{s}$ and $\boldsymbol{s} + 1$. Such anomalies occur because some path-following algorithms may employ occasional small predictor-corrector steps that are at least an order of magnitude smaller than standard steps. In this analysis it is observed that such intermittent relatively small steps in turn cause very small *BCP* shift vectors $\mathbf{dr}$ to be interspersed between longer runs of larger changes, causing 'spike' noise in the otherwise smooth eigenvector-space trajectories $\mathbf{T}(s)$. Such 'spikes', which usually only consist of a single spurious point deviating from the locally smooth eigenvector-space trajectory, can make potentially large spurious contributions to the eigenvector-space trajectory $\mathbf{T}(s)$ and may be safely filtered. A combination of criteria are recommended for automated rejection or inclusion of a specific point into the eigenvector-space trajectories $\mathbf{T}(s)$:

1. If the magnitude of the *BCP* shift vector $\mathbf{dr}$ associated with any current $\mathbf{T}(s)$ point is less than 50% of the average of the corresponding *BCP* shift vector $\mathbf{dr}$ values associated with the immediately preceding point and the immediately following point, the current point is filtered out as a 'spike'.
2. Abrupt changes in direction in the $\mathbf{T}(s)$, e.g. turning by more than $60^{o}$ from one $\mathbf{T}(s)$ step to the next cause the current point to be labelled as a 'spike'.

These two rules taken together are referred to as the 'turn' filter. These rules can be repeatedly applied across multiple 'passes' through the eigenvector-space trajectory **T**(*s*) data as necessary. Finally, a three-pass two-point averaging Kolmogorov-Zurbenko data filter, see reference 38 in the main text, is applied to the eigenvector-space trajectories **T**(*s*). These calculations of the eigenvector-space trajectories **T**(s) are implemented in our QuantVec program 'drproject3', see the **computational details** section of the main text.

**2(b)** *Quantities derived from the eigenvector-space trajectories **T(s)**.* The maximum range of eigenvector projections $\mathbf{T}(s)_{\mathbf{max}}$ = {bond-flexing$_{\mathbf{max}}$, bond-chirality$_{\mathbf{max}}$, bond-axiality$_{\mathbf{max}}$} are used to define the dimensions of a 'bounding box' around each **T**(*s*) and are used to calculate the bond-flexing **F**, bond-chirality **C** and bond-axiality **A** using the QuantVec program 'trajplot', see the **computational details** section of the main text. The subscript '$_{\mathbf{max}}$' corresponds to the difference between the minimum and maximum value of the projection of the *BCP* shift vector **dr** vector onto $\underline{\mathbf{e}}_1$, $\underline{\mathbf{e}}_2$ or $\underline{\mathbf{e}}_3$ along the entire **T**(*s*). Note: the $\underline{\mathbf{e}}_2$ corresponds to the direction in which the electrons at the *BCP* are subject to the most compressive forces, therefore $\underline{\mathbf{e}}_2$ corresponds to the most *facile direction* ('easy') for displacement of the *BCP* electrons. Conversely the $\underline{\mathbf{e}}_1$ ('hard') and $\underline{\mathbf{e}}_3$ correspond to the directions associated with the least compressive forces and the tensile forces on the *BCP* electrons respectively.

**Supplementary Materials S3.** Octopus input files for geometry optimization, ground and excited state preparation and propagation of electron and nuclear dynamics and specifications of the simulation box.

The simulation box for the real-space calculations was defined as a cuboid aligned along the ***x-y-z*** orthogonal Cartesian axes, with dimensions 8.0Å × 8.0Å ×18.0Å respectively. The centre of the box was placed at the Cartesian axes origin. The initial linear molecular structure was then aligned on the *z*-axis at the centre of the simulation box, see **Scheme 1**. The size of the simulation box was chosen as large enough, later confirmed after initial total electronic charge density $\rho(\mathbf{r})$ calculation, to ensure that the $\rho(\mathbf{r})$ at the edges of the box and hence potential boundary wave-packet reflection effects are negligible. The mesh spacing for the simulation box real-space grid was selected to be 0.05Å, a spacing known from previous work to provide well-converged QTAIM and NG-QTAIM properties[23,42].

**Geometry optimization**

```
#  HCCI molecule, initial geometry
# Default to atomic units
# Structure from Kraus et. al. 10.1126/science.aab2160 (SM 5), CCSD(T)
# cc-pVQZ/ECP28MDF
CalculationMode = go
FromScratch = yes
ExperimentalFeatures = true

# Geometry optimization
GOMethod = fire
GOTolerance = 0.01*eV/angstrom

# Simulation box and geometry
Dimensions = 3
PeriodicDimensions = 0
Spacing = 0.05*angstrom
BoxShape = parallelepiped
%Lsize
4.0 | 4.0 | 9.0
%
%BoxCenter
0.0000 | 0.0000 | -4.5000
%
# Atomic masses
# H 1.00784
# C 12.011
# I 126.90447
%Coordinates
"H" | 0.000000 | 0.000000 | -7.247245 | yes
"C" | 0.000000 | 0.000000 | -5.236745 | yes
"C" | 0.000000 | 0.000000 | -2.949445 | yes
"I" | 0.000000 | 0.000000 |  0.832255 | yes
%
SpinComponents = unpolarized
PseudopotentialSet = hgh_lda
ConvRelDens=1.0e-9
EigensolverTolerance=1.0e-8
Eigensolver=chebyshev_filter
MaximumIter=-1
ExtraStates = 8
ExtraStatesToConverge = 8
```

**Convergence of ground state**

```
CalculationMode = gs
FromScratch = yes
ExperimentalFeatures = true
# Simulation box and geometry
Dimensions = 3
PeriodicDimensions = 0
Spacing = 0.05*angstrom
BoxShape = parallelepiped
%Lsize
4.0 | 4.0 | 9.0
%
%BoxCenter
0.0000 | 0.0000 | -4.5000
%
# Atomic masses
# H 1.00784
# C 12.011
# I 126.90447
# New coordinates after Geometry optimization at HGH_LDA
%Coordinates
"H" | 0.000000 | 0.000000 | -7.003370583 | yes
"C" | 0.000000 | 0.000000 | -5.015085 | yes
"C" | 0.000000 | 0.000000 | -2.772519658 | yes
"I" | 0.000000 | 0.000000 |  0.792684423 | yes
%
SpinComponents = unpolarized
PseudopotentialSet = hgh_lda
ConvRelDens=1.0e-9
EigensolverTolerance=1.0e-8
Eigensolver=chebyshev_filter
MaximumIter=-1
ExtraStates = 8
ExtraStatesToConverge = 8
```

**Convergence of (initially unoccupied) excited states**

```
CalculationMode = unocc
FromScratch = yes
ExperimentalFeatures = true
# Simulation box and geometry
Dimensions = 3
PeriodicDimensions = 0
Spacing = 0.05*angstrom
BoxShape = parallelepiped
%Lsize
4.0 | 4.0 | 9.0
%
%BoxCenter
0.0000 | 0.0000 | -4.5000
%
# Atomic masses
# H 1.00784
# C 12.011
# I 126.90447
# New coordinates after Geometry optimization at HGH_LDA
%Coordinates
"H" | 0.000000 | 0.000000 | -7.003370583 | yes
"C" | 0.000000 | 0.000000 | -5.015085 | yes
"C" | 0.000000 | 0.000000 | -2.772519658 | yes
"I" | 0.000000 | 0.000000 |  0.792684423 | yes
%
```

```
SpinComponents = unpolarized
PseudopotentialSet = hgh_lda
ConvRelDens=1.0e-9
EigensolverTolerance=1.0e-8
Eigensolver=chebyshev_filter
MaximumIter=-1
ExtraStates = 8
ExtraStatesToConverge = 8
```

**Propagation of electron and nuclear dynamics (example for CW [+1] polarization)**

```
CalculationMode = td
FromScratch = no
ExperimentalFeatures = true
# Simulation box and geometry
Dimensions = 3
PeriodicDimensions = 0
Spacing = 0.05*angstrom
BoxShape = parallelepiped
%Lsize
4.0 | 4.0 | 9.0
%
%BoxCenter
0.0000 | 0.0000 | -4.5000
%
# New coordinates after Geometry optimization at HGH_LDA
%Coordinates
"H" | 0.000000 | 0.000000 | -7.003370583 | yes
"C" | 0.000000 | 0.000000 | -5.015085 | yes
"C" | 0.000000 | 0.000000 | -2.772519658 | yes
"I" | 0.000000 | 0.000000 |  0.792684423 | yes
%
SpinComponents = unpolarized
PseudopotentialSet = hgh_lda
ConvRelDens=1.0e-9
EigensolverTolerance=1.0e-8
Eigensolver=chebyshev_filter
MaximumIter=-1
ExtraStates = 8
ExtraStatesToConverge = 8
MoveIons = yes
MaximumIter=-1
#RestartWriteInterval = 50

# PROPAGATION
TDTimeStep = 0.01
TDEnergyUpdateIter = 1
TDPropagator = etrs
RecalculateGSDuringEvolution = yes
RecalculateGSInterval = 50
TDStepsWithSelfConsistency = all_steps

# total propagation time
TDPropagationTime = 25.0*fs
#TDMaxSteps = 1
TDSCFThreshold = 1.0e-8
TDExponentialMethod = lanczos

# PULSE PARAMETERS
# single sin^2 envelope circular polarized pulse in plane starting at t=0,
# duration 'pulsetime', x-y plane, peak field 'amp' (au), carrier freq 'freq' (au)
#
sqrt2=1.4142135623730950
```

```
freq=0.2057962049           # carrier frequency (au) = 5.60eV from SM(5)
amp=0.0100 # pulse max E-field amplitude (au, converted from Gv/m)
pulsetime=2.0        # duration of single pulse (fs)
pulsetime_au=pulsetime*fs  # duration of single pulse (au, computed)
cep=0.0              # relative phase of carrier and envelope (degrees)
cep_r=cep*pi/180.0       # relative phase of carrier and envelope (radians, computed)

# specify time-dependent fields
%TDExternalFields
    electric_field | 1 | i | 0 | freq | "envelope_sin2" | "cepfunc"
%
%TDFunctions
    "envelope_sin2"  |  tdf_from_expr   |  "sqrt2*amp*sin((pi*t/pulsetime_au))^2*(1-step(t-
pulsetime_au))"
    "cepfunc"       | tdf_from_expr | "cep_r"
%

OutputInterval=16
%Output
  geometry | "output_format" | xyz | "output_interval" | 16
  density | "output_format" | cube | "output_interval" | 16
%
TDOutputComputeInterval = 16
%TDOutput
  laser
  coordinates_sep
  velocities_sep
  forces_sep
%
```

**Supplementary Materials S4.** The variation of the eigenvector projections $t_1$, $t_2$ and $t_3$ with time of the iodoacetylene molecular graph.

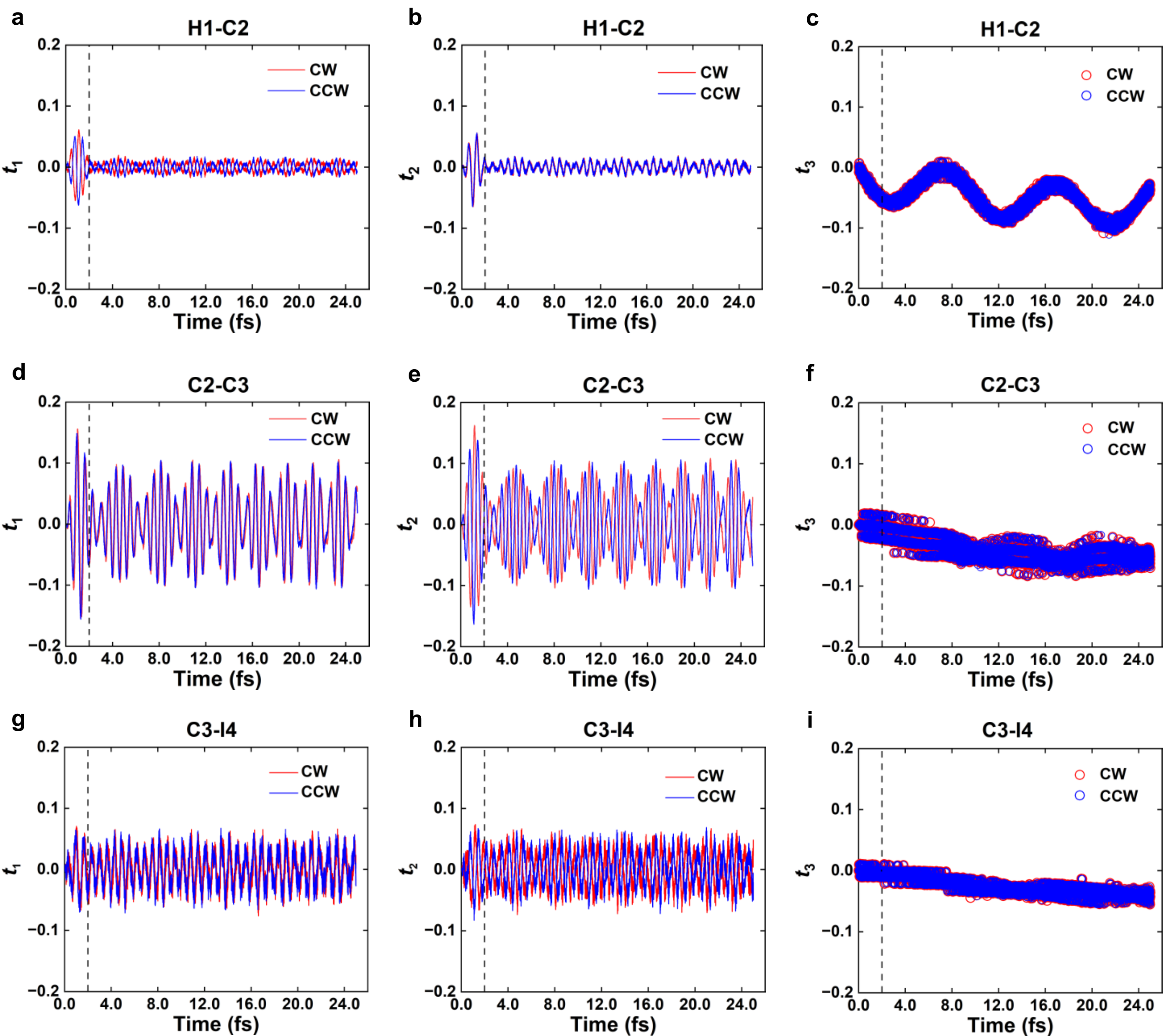

**Supplementary Figure S4**. The variation of eigenvector projections $t_1$ = **e₁·dr**, $t_2$ = **e₂·dr** and $t_3$ = **e₃·dr** for the H1-C2 *BCP* (upper panel), C2-C3 *BCP* (middle panel), and C3-I4 *BCP* (lower panel) with time (femtoseconds (fs)) of the iodoacetylene molecular graph for the {CW, [+1]} (red) and {CCW, [-1]} (blue). The plots are constructed with the *BCP* shift vector **dr** along with corresponding Hessian $\underline{\mathbf{e}}_1$, $\underline{\mathbf{e}}_2$ and $\underline{\mathbf{e}}_3$ eigenvectors at the *BCP* at time $t = 0$. Results are presented for the duration of the simulated laser pulses ($0 \leq t \leq 2.00$ fs) and the duration of the 'after pulse' ($2.00$ fs $\leq t \leq 23$ fs), that are separated by the dashed vertical line at 2.00 fs. The QuantVec 'drproject3' program is used to construct the eigenvector projections.

**Supplementary Materials S5.** Variation of the iodoacetylene scalar measures with time.

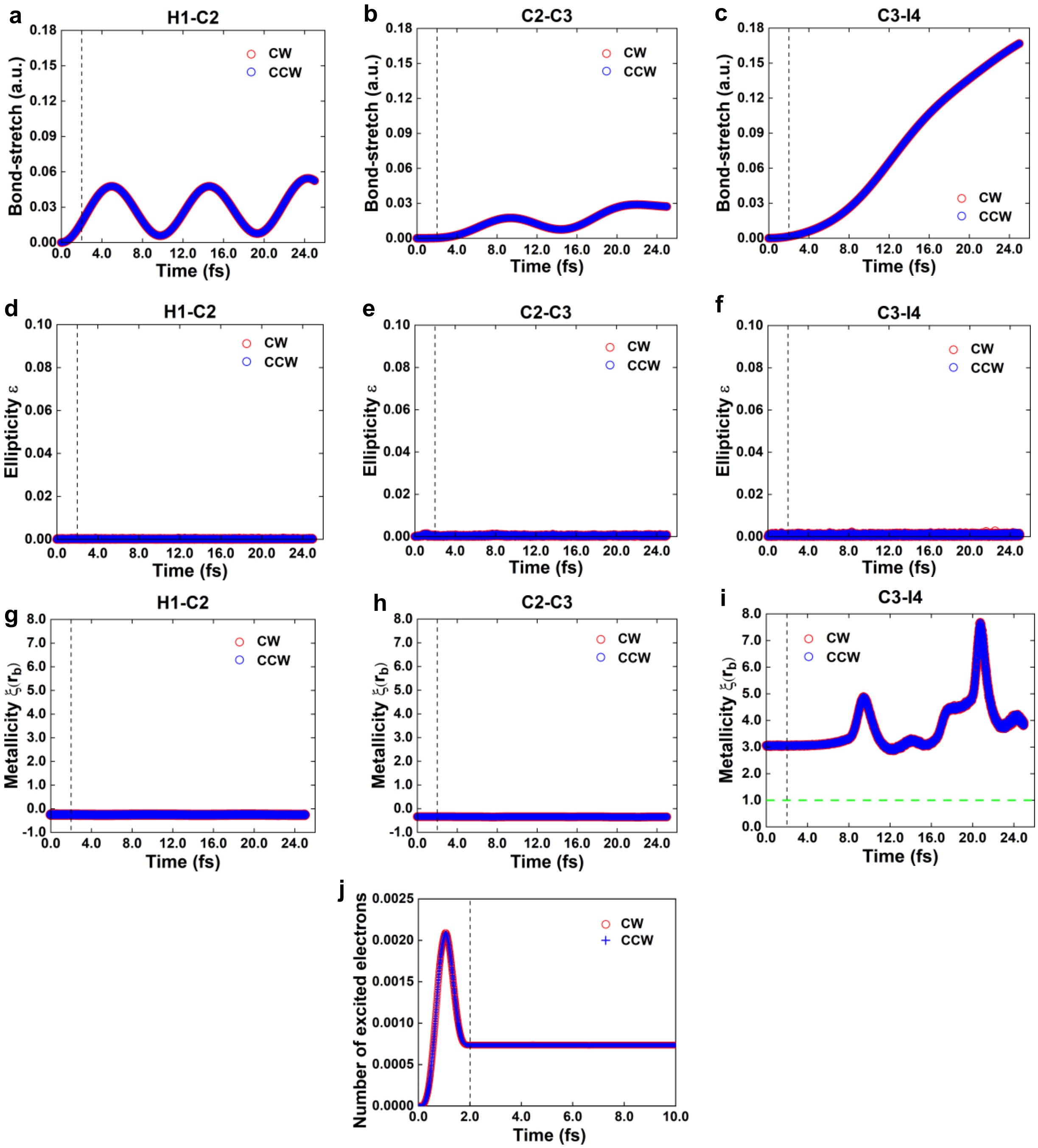

**Supplementary Figure S5**. The variations of the scalar quantities the iodoacetylene molecule with time of the clockwise {CW, [+1]}, right-handed (**R**, red) and left-handed, counter-clockwise {CCW, [-1]}, left-handed (**S**, blue) circularly polarized simulated laser pulses with a peak E-field value $E = 100.0 \times 10^{-4}$ a.u. is presented. The variations of the interatomic separation (GBL) minus the interatomic separation before the laser is applied ($GBL_0$), i.e. the bond-stretch ($GBL - GBL_0$), the $GBL_0$ values are 2.010 a.u, 2.287 a.u and 3.871 a.u respectively, are provided in sub-figure **(a-c)**. The corresponding variations of the iodoacetylene molecular graph ellipticity $\varepsilon$ for the H1-C2 *BCP* (left), C2-C3 *BCP* (middle) and C3-I4 *BCP* (right) are provided in sub-figure **(d-f)**. Results are presented for the duration of the pulse ($0 \le t \le 2.00$ fs) and the 'after pulse' ($2.00$ fs $\le t \le 23$ fs), the dashed vertical line at 2.0 fs indicates the end of the laser pulses. In the absence of an applied **E**-field there is no significant ellipticity $\varepsilon$ present for any of the *BCP*s. The corresponding variation of the iodoacetylene molecular graph metallicity $\xi(\mathbf{r}_b)$ is provided in sub-figure **(g-i)**, note the green dashed line at $\xi(\mathbf{r}_b) = 1.0$ indicating the lowest threshold for significant $\xi(\mathbf{r}_b)$ values. The variations of the numbers of excited electrons is provided in sub-figure **(j)**. The QuantVec 'cpscalar' program was used to calculate the bond-stretch ($GBL - GBL_0$), sub-figure **(a-c)**, ellipticity $\varepsilon$, sub-figure **(d-f)** and metallicity $\xi(\mathbf{r}_b)$, sub-figure **(g-i)**.

**Supplementary Materials S6.** Table of the *BCP* eigenvector-space trajectory $\mathbf{T}(s)_{max}$ bounding boxes of the iodoacetylene molecular graph.

The components $\{(\underline{\mathbf{e}}_1\cdot\mathbf{dr})_{max}, (\underline{\mathbf{e}}_2\cdot\mathbf{dr})_{max}, (\underline{\mathbf{e}}_3\cdot\mathbf{dr})_{max}\}$ of the *BCP* $\mathbf{T}(s)_{max}$ bounding boxes, the maximum range of the Hessian of $\rho(\mathbf{r})$ *BCP* eigenvector-space trajectory $\mathbf{T}(s)$ projections with simulated non-ionizing clockwise {CW, [+1]}, right-handed (**R**) and counter-clockwise {CCW, [-1]}, left-handed (**S**) circularly polarized laser pulses with peak electric (E)-field E = 100.0 x $10^{-4}$ a.u. All entries have been multiplied by $10^3$. Results are presented for the duration of the pulse ($0 \leq t \leq 2.00$ fs) and the duration of the 'after pulse' ($2.00$ fs $\leq t \leq 23$ fs), also see **Scheme 1** of the main text. The QuantVec 'drproject3' program is used to construct the *BCP* $\mathbf{T}(s)$. The QuantVec program 'trajplot' is used to calculate the components $\{(\underline{\mathbf{e}}_1\cdot\mathbf{dr})_{max}, (\underline{\mathbf{e}}_2\cdot\mathbf{dr})_{max}, (\underline{\mathbf{e}}_3\cdot\mathbf{dr})_{max}\}$ of each *BCP* $\mathbf{T}(s)_{max}$ bounding box.

| ***BCP T(s)*** | ***BCP T(s)$_{max}$ bounding boxes*** | |
|---|---|---|
| | **CW (R**, [+1]) | **CCW (S**, [-1]) |
| | $\{(\underline{\mathbf{e}}_1\cdot\mathbf{dr})_{max}, (\underline{\mathbf{e}}_2\cdot\mathbf{dr})_{max}, (\underline{\mathbf{e}}_3\cdot\mathbf{dr})_{max}\}$ | $\{(\underline{\mathbf{e}}_1\cdot\mathbf{dr})_{max}, (\underline{\mathbf{e}}_2\cdot\mathbf{dr})_{max}, (\underline{\mathbf{e}}_3\cdot\mathbf{dr})_{max}\}$ |
| *During pulse* | | |
| *H1-C2* | {0.113388, 0.116550, 0.055715} | {0.111618, 0.116260, 0.055784} |
| *C2-C3* | {0.302651, 0.293911, 0.032992} | {0.301049, 0.298360, 0.032752} |
| *C3-I4* | {0.125775, 0.125122, 0.016162} | {0.116630, 0.130417, 0.015911} |
| *After pulse* | | |
| *H1-C2* | {0.031999, 0.033075, 0.101178} | {0.030957, 0.031770, 0.100641} |
| *C2-C3* | {0.204305, 0.210819, 0.071554} | {0.206461, 0.209996, 0.071914} |
| *C3-I4* | {0.117930, 0.119016, 0.047886} | {0.119127, 0.117076, 0.048142} |